\documentstyle[spie]{article} 
\input epsf.sty
\begin{document}
\title{Role of Symmetry in Raman Spectroscopy of Unconventional 
Superconductors}
\author{T. P. Devereaux \\[12pt]
 Department of Physics\\ 
University of California, Davis, CA 95616\\[12pt]}
\maketitle
\date{\today}
\begin{abstract}
The role of symmetry of the inelastic light scattering amplitude,
the superconducting energy gap, and the underlying Fermi surface
manifold on the Raman spectra of unconventional superconductors is 
discussed in detail. Particular emphasis is placed on both
single and bi-layer superconductors. It is found that the $B_{1g}$
channel may be the most sensitive to doping due to the role of the Van Hove
singularity. Lastly the effect of both disorder and spin fluctuations is 
considered. The theory imposes
strong constraints on both the magnitude and symmetry of the energy gap 
for the bi-layer cuprates, indicating that a nearly identical energy gap
of $d_{x^{2}-y^{2}}$ symmetry provides a best fit to the data. \\
\centerline{Keywords: Raman scattering, unconventional superconductivity, bi-layer, spin fluctuations, impurities.} 
\end{abstract}
\section{Introduction}
In the recent few years Raman scattering has been proved to be a very
powerful tool to study the anisotropy of the superconducting energy gap
in high T$_{c}$ superconductors. The rich information
provided from experiments lies in the polarization dependence of the
obtained spectra for small light energy transfers. The recent experiments
carried out in different high T$_{c}$ materials near optimal doping
provide almost
identical spectra showing very characteristic differences between
the various polarization orientations\cite{tpd,irwin,exp}. This polarization 
dependence, which is minimal in conventional superconductors, can be 
understood in unconventional superconductors primarily in terms of
symmetry.

The light scattering amplitude can be given to lowest order in the
vector potential as a sum of the three terms which
correspond to direct photon scattering via electron-hole pair production and
via photon absorption and emission through an intermediate state.
In the limit
of zero momentum photon transfer this vertex can be written in the 
form\cite{kandd,ag}
\begin{equation}
\gamma({\bf k};\omega_{I},\omega_{S})= {\bf e}^{I}\cdot {\bf e}^{S}
+m^{-1}\sum_{\nu}\left\{
{({\bf k}\mid {\bf k}\cdot {\bf e}^{S}\mid \nu {\bf k})
(\nu {\bf k}\mid {\bf k}\cdot {\bf e}^{I}\mid {\bf k})\over{
\epsilon_{\bf k}-\epsilon_{\nu,{\bf k}}+\omega_{I}}} +
{({\bf k}\mid {\bf k}\cdot {\bf e}^{I}\mid \nu {\bf k})
(\nu {\bf k}\mid {\bf k}\cdot {\bf e}^{S}\mid {\bf k})\over{
\epsilon_{\bf k}-\epsilon_{\nu,{\bf k}}-\omega_{S}}}\right\},
\end{equation}
where $m$ is the electron mass, $\nu$ stands for the intermediate state
$\mid \nu {\bf k}\rangle$
of the electron excited out of the conduction band $\mid {\bf k}\rangle$ with
dispersion $\epsilon_{\bf k}$, $\omega_{I,S}$ and ${\bf e}^{I,S}$
are the incident and scattered photon energies and polarizations,
respectively. 

According to Abrikosov and Genkin if the energy of the incident
and scattered frequencies are negligible compared to the relevant 
electronic energy scale, $\gamma$ can be expressed in terms 
of the curvature of the conduction band and the incident
and scattered photon polarization vectors ${\bf e}^{I,S}$ as\cite{ag}
\begin{equation}
\lim_{\omega_{I},\omega_{S} \rightarrow 0}\gamma({\bf k};\omega_{I},\omega_{S})
=\sum_{\mu,\nu} e^{I}_{\mu}{\partial^{2}\epsilon({\bf k})\over{\partial k_{\mu}
\partial k_{\nu}}}e^{S}_{\nu},
\end{equation}
where terms of the order of $1-\omega_{S}/\omega_{I}$ are dropped.
This effective mass approximation has been extensively used to calculate the 
Raman spectra in unconventional superconductors. 

This approximation holds only for small energy transfers {\it and} provided
that the bands are not nearly degenerate. Thus one might question the 
appropriateness of this approximation for the cuprates given
that typical incoming laser frequencies are on the order of 2 eV -  certainly
on the order of 
the relevant electronic energy scale for a single band. 
For the case of two bands near or crossing the Fermi level, the
effective mass approximation is even more questionable.
Since in the bi-layer superconductors Bi 2:2:1:2
and Y 1:2:3 the band splitting of the even and odd bands is on the order
of tens of meVs, the approximation misses large terms corresponding to
interband transitions at relatively small energies.
Detailed knowledge of the magnitude of the scattering amplitude in this
case requires knowledge of the wave functions of the two states, which has
currently not been investigated. Thereby, use of the
effective mass approximation is uncontrolled in many band systems and 
comparisons to experiment intensities must be viewed with caution.

An alternative approach is based on the experimental observation that
the spectra near optimal doping for a
wide range of cuprate materials depends only mildly on the incoming laser
frequency. Since the polarization orientations transform as various
elements of the point group of the crystal, one can use symmetry to classify
the scattering amplitude, viz.,
\begin{equation}
\gamma({\bf k};\omega_{I},\omega_{S})=\sum_{L}\gamma_{L}(\omega_{I},\omega_{S})
\Phi_{L}({\bf k}),
\end{equation}
where $\Phi_{L}({\bf k})$ are either Brillouin zone (B.Z.H., orthogonal 
over the entire Brillouin zone) or Fermi surface (F.S.H., orthogonal
on the Fermi surface only) harmonics which transform
according to point group transformations of the crystal\cite{allen}. 
Representing the magnitude but not the 
${\bf k}$-dependence of both intra- and interband scattering, the prefactors
can be approximated to be frequency independent and taken as model constants 
to fit absolute intensities. 
Thus we have simplified the many-band problem in terms
of symmetry components which can be related to charge degrees of freedom
on portions of the Fermi sheets. While sacrificing information
pertaining to overall intensities, we have gained the ability to probe
and compare excitations on different regions of the Fermi surface based
solely on symmetry classifications. This can be illustrated
by considering the various experimentally accessible polarization orientations.

In the following we confine ourselves to tetragonal systems with in-plane
lattice constant $a$, and for the moment consider uncoupled planes which
can be modelled by the following band structure:
\begin{equation}
\epsilon({\bf k})=-2t[\cos(k_{x}a)+\cos(k_{y}a)]+4t^{\prime}
\cos(k_{x}a)\cos(k_{y}a)-2t^{\prime\prime}[\cos(2k_{x}a)+\cos(2k_{y}a)]
-\mu.
\end{equation}
We will consider coupling between the planes in the second part of
the following section.
Using an $x,y$ coordinate system locked to the CuO$_{2}$ planes, incident and 
scattered light polarizations aligned
along $\hat x+\hat y, \hat x-\hat y$ for example
transform according to $B_{1g}$ symmetry, and thus 
\begin{equation}
\Phi_{B_{1g}}({\bf k})=\cos(k_{x}a)-\cos(k_{y}a) + \dots,
\end{equation}
where $\dots$ are higher order B.Z.H.
Likewise, ${\bf e}^{I,S}$ aligned
along $\hat x, \hat y$ transforms as $B_{2g}$:
\begin{equation}
\Phi_{B_{2g}}({\bf k})= \sin(k_{x}a)\sin(k_{y}a) + \dots.
\end{equation}
The $A_{1g}$ basis function is
\begin{equation}
\Phi_{A_{1g}}({\bf k})=
a_{0} + a_{2}[\cos(k_{x}a)+\cos(k_{y}a)] + a_{4}\cos(k_{x}a)\cos(k_{y}a)
+a_{6}[\cos(2k_{x}a)+\cos(2k_{y}a)] +\dots, 
\end{equation}
where the expansion parameters $a_{i}$ determined via a fitting procedure
with experiment\cite{efm}. The $A_{1g}$ response is not directly accesible
from experiments and must be obtained by subtracting several combinations
of the response for various polarization orientations. 

By considering the ${\bf k}-$dependence of the basis functions, it is clear
that the $B_{1g}$ part of the spectra essentially probes light scattering
events along the $k_{x}$ or $k_{y}$ axes, $B_{2g}$ probes the diagonals,
and $A_{1g}$ is more of an average over the entire Brillouin zone. It is in 
this manner that information about the momentum dependence of the 
superconducting energy gap can be usefully extracted from the data\cite{tpd}.
That information lies in two important aspects: the low frequency power-
law behavior of the spectra and the positions of the low energy peaks.

\section{Raman Response for Unconventional Superconductors}
The spectra can be obtained via the two-particle Raman correlation function
and is derivable either via Green's function or kinetic equation approach.
In the following sections, the response is calculated for the case of a 
single band model and a superconducting bi-layer. 

\subsection{Single Band}

For a single band crossing the Fermi surface, the response can be written
in the ``Pair Approximation'' \cite{prange} in terms of the Tsuneto function
$\lambda$ as\cite{tpdde}
\begin{equation}
\chi_{\gamma,\gamma}({\bf q}=0,i\omega)=\sum_{\bf k} \gamma({\bf k};
i\omega_{I},i\omega_{S})^{2}
\lambda({\bf k},i\omega);\ \ \ 
\lambda({\bf k}, i\omega)=
{\Delta({\bf k})^{2}\over{E({\bf k})^2}}
\tanh\left[{E({\bf k})\over{2 T}}\right]
\left[{1\over{2E({\bf k})+i\omega}}+
{1\over{2E({\bf k})-i\omega}}\right].
\end{equation}
Here $E({\bf k})^{2}=\epsilon({\bf k})^{2}
+\Delta({\bf k})^{2}$ and we have set $k_{B}=\hbar=1$. The light scattering
cross section is obtained by taking the imaginary part via analytic
continuation. We have neglected 
vertex corrections of the pairing interactions which while responsible for 
maintaining gauge invariance and producing collective modes, can be shown
to have only a limited effect on the spectra at low energies for $d-$wave
superconductors\cite{tpdde}. The long-range part of Coulomb
interaction is not included at this stage, but will be shown to be very 
important for the $A_{1g}$ response.

Eq. (8) and approximations to it have been evaluated previously for
various forms of the energy gap $\Delta({\bf k})$. The response has been
evaluated using a simpler tight binding parametrization of the energy band
dispersion by (i) restricting the sum to the Fermi surface using a
weak coupling energy gap\cite{irwin,tpdde}, 
and (ii) performing the ${\bf k}-$ sums directly while using
a phenomenological spin fluctuation mediated energy gap\cite{carbotte}.
These are improvements on the original consideration which approximated the
Fermi surface to be nearly cylindrical\cite{tpd}. 

In this section the $T << T_{c}$ Raman response for a single CuO layer
is obtained by evaluating the ${\bf k}$-sums directly
in order to capture the charge dynamics over the entire Brillouin zone, and
in particular the role of the Van Hove singularities can be examined.  
We focus attention on La 2:1:4\cite{levin} and Y 1:2:3\cite{oka} 
as representive calculations. 

The $B_{1g}$ and $B_{2g}$ responses are plotted in Fig. 1 using an energy
gap of $d_{x^{2}-y^{2}}$ symmetry. In both cases, the $B_{1g}$ response for low
frequencies varies as $\omega^{3}$ while the $B_{2g}$ varies linearly
with $\omega$. This can be shown to be generally true for these channels
for any number of higher harmonics used, and results from a
consideration of the density of states (DOS), which varies 
linearly with energy due to the nodes, and the behavior of the Raman 
vertex $\gamma$ near the 
gap nodes. Since the $B_{2g}$ vertex is finite near the nodes, the DOS 
determines the low frequency behavior. However, the gap and the vertex vanish 
at the same place for the $B_{1g}$ channel, which in turn leads to the cubic 
behavior. This delicate interplay of vertex and energy gap can thus provide
a unique determination of the nodal gap behavior.

Moreover, additional information lies in the peaks of the spectra. A smooth
peak is seen for $B_{2g}$ and a double-peak is obtained for $B_{1g}$. The
lower peak is due to the energy gap while the other comes from the Van
Hove singularity. The peak of the spectra due to superconductivity in both
cases is higher for the 
$B_{1g}$ channel ($\omega_{peak}=2\Delta_{max}$)
than for $B_{2g}$, ($\omega_{peak}\sim 1.7\Delta_{max}$) since
the gap maxima are located at the
same part of the Fermi surface where the $B_{1g}$ basis function is
largest, and where the $B_{2g}$ basis function vanishes. This is generally
true regardless of Fermi surface shape, although the relative positions of
the peaks can be affected slightly by changing the underlying manifold 
(see Fig 2a and 2b and Refs. \cite{tpd,irwin,tpdde}). 
The shape of the spectra can also by mildly modified 
by considering higher
harmonics of the energy gap \cite{carbotte} and/or by including final-state
interactions\cite{tpdde} or impurity scattering\cite{imp1}.
Apart from the last case, the power-law behaviors at low frequencies
are unaffected and therefore their observation is a robust check on the
symmetry of the energy gap. 

The Van Hove peak only shows up in the $B_{1g}$ channel since the 
$B_{2g}$ channel assigns no weight to the location of the Van Hove points. 
The Van Hove peak lies at
a higher energy determined by its location off the Fermi surface and the
value of the energy gap there. There are indications
that a double-peak feature has been seen in underdoped Y 1:2:3
with a T$_{c}=60 K$\cite{slakey}. As the Van Hove moves further away from the
Fermi surface for example with doping, the peak shifts to higher 
frequencies and has a smaller
residue. Inelastic scattering at higher energies will also act to smear
out this feature and thus a single smeared and perhaps asymmetric peak
would be seen if the Van Hove lies quite near the Fermi surface, and could
in principle move the $B_{1g}$ peak slightly above $2\Delta_{max}$. This
may be reflected in the experiments of Ref. \cite{alt} which saw a
sensitivity of the $B_{1g}$ peak position to doping.

We now turn to the $A_{1g}$ channel. Here the long-range Coulomb screening 
plays a very important role. Namely, those parts of the spectra which are
coupled to the Coulomb fields are screened and do not appear in the 
limit of small momemtum transfer $q \rightarrow 0$, as is always
the case. The long-range Coulomb forces play an important role if
there is charge transfer between unit cells at large distances. Thus
only the charges produced by charge transfers inside the unit cell
determine the measurable spectra. This transfer is between the different
atoms in the cell or the redistribution of the electrons between
different parts of the Fermi surface or between different Fermi surfaces.
In this way, only the $A_{1g}$ part of the spectra can be coupled to the
Coulomb forces since for the case of $B_{1g}$ and $B_{2g}$
channels, the Coulomb corrections vanish due to symmetry.

Including the long-range Coulomb
interaction, the $A_{1g}$ response can be written as\cite{kandd,tpdde}
\begin{equation}
\chi^{A_{1g}}_{\gamma,\gamma}(i\omega)=\chi_{\gamma,\gamma}(i\omega)
-{[\chi_{1,\gamma}(i\omega)]^{2}\over{\chi_{1,1}(i\omega)}},
\end{equation}
where $\chi_{a,b}$ is given by Eq. (8) with vertices $a,b$ replacing
$\gamma$. Writing the $A_{1g}$ vertex as $\gamma({\bf k})=
\langle \gamma({\bf k}) \rangle +\delta\gamma({\bf k})$, where
$\langle \gamma({\bf k}) \rangle$ is the average vertex over the Brillouin
zone, and inserting into Eq. (9) one finds that indeed
the long-range Coulomb interaction completely
screens the constant contribution to the vertex, or the intercell
charge fluctuation contribution, and thus only the intracell fluctuations
can scatter light in the limit of $q=0$. 

In Fig. 2 we show a calculation of
the unscreened and screened $A_{1g}$ response for a 
$d_{x^{2}-y^{2}}$ paired superconductor for the Y 1:2:3 parameters. We
have chosen the coefficients $a_{i}$ in Eq. (7) using the effective mass
approximation\cite{efm}, but we note that the results do not depend
dramatically on the choice of coefficient sets. By again considering the
nature of the nodal structure of the energy gap, since the $A_{1g}$ response
measures a weighted average of charge fluctuations around the Fermi surface,
it can be shown that the low energy part of the $A_{1g}$ response must vary 
with energy in the same way as the DOS. 

The screened and unscreened responses at higher energies are very different. 
For the unscreened response, two peaks arise 
corresponding to the singular contribution of pair breaking
along the axes and the Van Hove contribution as in the $B_{1g}$ channel. 
Screening completely 
reorganizes the spectra to remove both peak contributions, replacing
them with a much smoother peak at lower energies. This is due to the
screening function (2nd part of Eq. (9)) which also contains 
peaks at $2\Delta_{max}$ due to the weighting along the $k_{x}$ and $k_{y}$
axes, and the Van Hove. These peaks cancel the unscreened peaks and creates a 
near peak lying at slightly greater than $\Delta_{max}$, as seen in Fig. 2. 

In this manner, similar symmetry considerations can be made for various 
types of energy gaps. The Raman spectra calculated for various energy gaps
is summarized in Table 1 and compared to the low energy DOS $N(\omega)$.
The dominant contribution to the Raman lineshape
is due to the location and behavior of the energy gap near both the nodes and 
the maxima. Subsequently, different energy gaps produce different line
shapes as summarized in Table 1. Here it is important to note that the
cubic rise of the spectra in any channel requires that both the vertex
and the energy gap have the same behavior near the nodes. In 
tetragonal systems, this requires the presence of a $d_{x^{2}-y^{2}}$ energy 
gap. By considering small orthorhombic distortions, the $A_{1g}$ and $B_{1g}$ 
channels become
mixed and therefore a linear rise with frequency could be obtained at low
energies in a region determined by the amount of symmetry breaking.
Since experimentally
the low frequency part of the $B_{1g}$ spectra rises cubically in the most
tetragonal systems, while in Y 1:2:3 the spectra has a small linear part,
this strongly suggests the likelihood of a $d_{x^{2}-y^{2}}$ pair state 
for these systems. Exploration of the low
frequency part of the spectra could put stringent constraints on other
pair-state candidates.

\subsection{Bi-Layer}

The above consideration must be reformulated for the case of two or more
bands crossing the Fermi level\cite{dvz}. In particular, the role of 
screening is
more subtle in this case. The above arguments on the role of Coulomb
screening is based on the assumption that only one
Fermi surface is relevant, which is very reasonable for materials with
a single CuO$_{2}$ plane in the unit cell. Recently Krantz and Cardona
\cite{candk} have raised the relevant question how a double-sheeted Fermi
surface, as occurs in materials with more than one CuO$_{2}$ planes in
the unit cell, changes the above argument. In particular, the main question 
is whether or not in the bi-layer compound a singularity
can occur at 2$\Delta_{max}$ in the $A_{1g}$ symmetry, similar to the
one which occurs in the $B_{1g}$ case, due to the avoidance of screening. 

Considering the
electronic density, the singular regions of the Fermi surface are
near the $k_{x}$ and $k_{y}$ axes, respectively. The two layers
double the number of these regions. In the case of $B_{1g}$ symmetry
the charge transfer changes sign going from one region to the other.
In contrast, in the case of $A_{1g}$ symmetry, each region has the same
phase on a particular Fermi surface. Therefore, in the case of the 
single layer material, considering the singular part, the charge
transfer is only between different cells, which is screened. In the 
bi-layer case, the $A_{1g}$ symmetry allows a charge transfer 
between the two layers of both the same and opposite sign on each
Fermi surface. While the addition of the two charge density distributions
is still an intercell charge transfer and is thus screened, if the bands are
non-degenerate then the difference of the two distributions can survive
screening and could thus in principle give a singular contribution at
$2\Delta_{max}$.

It is straightforward repeat the previous calculations for the two band
model. The calculation proceeds by first considering charge density-like
fluctuations on each CuO$_{2}$ plane and allow for charge to be transferred
between the planes due to an interplane coupling matrix element $t_{\perp}$
through which electrons can hop from one plane to the other directly or 
through an intermediate state such as the chains or Y atoms. Details of the
calculation can be found in Ref. \cite{dvz}.
Diagonalizing the Hamiltonian by introducing even and odd combinations
of electron operators in the two planes, we arrive at the following compact
expression for the cross section:
\begin{equation}
\chi^{\prime\prime}(\omega)=
\sum_{{\bf k},\pm}[\gamma({\bf k})\pm\gamma_{1,2}({\bf k})]^{2}
\lambda_{\pm}^{\prime\prime}({\bf k},\omega) 
-\left\{{\left( \sum_{{\bf k},\pm}[\gamma({\bf k})\pm\gamma_{1,2}({\bf k})]
\lambda_{\pm}({\bf k},i\omega)\right)^{2}\over{\sum_{{\bf k},\pm}
\lambda_{\pm}({\bf k},i\omega)}}\right\}^{\prime\prime}. 
\end{equation}
Here $\lambda_{\pm}$ is the Tsuneto function for the bonding (anti-bonding)
band, $\epsilon_{\pm}({\bf k})=\epsilon({\bf k})\mp t_{\perp}({\bf k})$, 
and the vertices are given by
\begin{equation}
\gamma({\bf k})=\gamma_{1,1}({\bf k})+\gamma_{2.2}({\bf k})=
{1\over{2}}[\gamma_{+}({\bf k})+\gamma_{-}({\bf k})]; \ \ \ \ \
\gamma_{1,2}({\bf k})={1\over{2}}[\gamma_{+}({\bf k})-\gamma_{-}(\bf k)].
\end{equation}
This corresponds to a diagonal contribution which allows light to be
scattered by density-like
fluctuations on either plane 1 ($\gamma_{1,1}$) or plane
2 ($\gamma_{2,2}$) and an off-diagonal term ($\gamma_{1,2}$) which allows
light scattering on both planes simultaneously. Since in the
limit of ${\bf q} \rightarrow 0$ the long-range intercell fluctuations
are screened, only the intracell fluctuations remain. Therefore only
charge transfer fluctuations between the atoms in the plane and outside
the plane (e.g., $Y$ and the chains), and between the planes can effectively
cause light scattering. 
Since experimentally the interlayer coupling is small, the resulting
vertex $\gamma_{1,2}$ must be smaller than the $+$ combination, labelled
as $\gamma$. 

Rearranging terms, we can cast the result in terms of an addition
of the result for each single band plus a mixing term:
\begin{equation}
\chi^{\prime\prime}(\omega)=\chi^{\prime\prime}_{+}(\omega)+
\chi^{\prime\prime}_{-}(\omega)+\Delta\chi^{\prime\prime}(\omega), 
\end{equation}
where
\begin{eqnarray}
&&\chi^{\prime\prime}_{\pm}(\omega)=\sum_{\bf k}\gamma_{\pm}^{2}({\bf k})
\lambda_{\pm}^{\prime\prime}({\bf k},\omega) 
-\left\{{\left(\sum_{\bf k}\gamma_{\pm}({\bf k})\lambda_{\pm}({\bf k},i\omega)\right)^{2}
\over{\sum_{\bf k}\lambda_{\pm}({\bf k},i\omega)}}\right\}^{\prime\prime}; 
 \\
&&\Delta\chi^{\prime\prime}(\omega)= \left\{{\left(
\sum_{\bf k}\lambda_{+}({\bf k},i\omega)\right)\left(\sum_{\bf k}\lambda_{-}
({\bf k},i\omega)\right)\over{\sum_{{\bf k},\pm}
\lambda_{\pm}({\bf k},i\omega)}}\left[
{\sum_{\bf k}\gamma_{+}({\bf k})
\lambda_{+}({\bf k},i\omega)\over{\sum_{\bf k}\lambda_{+}({\bf k},i\omega)}}
-{\sum_{\bf k}\gamma_{-}({\bf k})
\lambda_{-}({\bf k},i\omega)\over{\sum_{\bf k}\lambda_{-}({\bf k},i\omega)}}
\right]^{2}\right\}^{\prime\prime}. 
\end{eqnarray}

Eqs. (12-14) shows that the total response can be
considered as a sum of the single band contributions in addition to
a mixing term which corresponds to odd combinations of fluctuations
on the bands simultaneously. In order to discern the
features of the mixing term, it is useful to write $\gamma_{\pm}({\bf k})
=\gamma_{\pm}^{0}+\Delta\gamma_{\pm}({\bf k})$, where the average of
$\Delta\gamma({\bf k})$ around the Fermi surface vanishes. 

Then it useful to consider the following cases: 

(1) If the layers are uncoupled, then $\gamma_{+}({\bf k})=\gamma_{-}
({\bf k})$ and the mixing term vanishes: $\Delta\chi^{\prime\prime}
(\omega)=0$.

(2) If the scattering in the $A_{1g}$ symmetry channel $\gamma$ is
independent of ${\bf k}$ (scattering on real charge), then
$\Delta\chi \sim (\gamma_{+}^{0}-\gamma_{-}^{0})^{2} \sim \gamma_{1,2}^{2}$.
This means that the light scattering induces an interlayer charge transfer.
However, it can be seen by the first term on the r.h.s. of Eq. (14) that 
in order for
a divergence to occur above the peak for the single band result, then the
energy gaps on each band must be identical. Otherwise, the divergences
in the numerator and denominator cancel each other and a smoothly
varying response is obtained.

(3) Considering when the light induces charge transfers on different
parts of the Fermi surfaces, $\Delta\gamma_{\pm}({\bf k})$ must be
included and includes different symmetry channels depending on the
orientation of the incident and scattered light polarizations. $A_{1g}$
symmetry reflects the approximate tetragonal symmetry of the CuO$_{2}$
planes, while the part of $\gamma$ showing symmetries different than
$A_{1g}$ (e.g., $B_{1g}, B_{2g}$ and $E_{g}$) changes sign when
symmetry transformations are applied. The Tsuneto function depends only
on the energy gap squared, and thus all the sums in Eq. (14)
{\it are zero except for the $A_{1g}$ component of $\Delta\gamma_{\pm}
({\bf k})$}. For tetragonal materials, this means that the $B_{1g},
B_{2g},$ and $E_{g}$ channels do not have a mixing term and therefore
the total response for these channels is simply the additive
contributions from each single band. The mixing term only contributes
to the $A_{1g}$ channel, the strength of which is roughly determined
by the interlayer coupling. 

These points can be easily illustrated by considering a bi-layer superconductor
with  energy gaps $\Delta_{\pm}({\bf k})=\Delta_{\pm}
[\cos(k_{x}a)-\cos(k_{y}a)]$ on the two bands. 
For the interlayer coupling
$t_{\perp}$, we follow Ref. \cite{oka} and define
\begin{equation}
t_{\perp}({\bf k})=t_{\perp}[\cos(k_{x}a)-\cos(k_{y}a)]^{2},
\end{equation}
with $t_{\perp}/t=0.2$. In the absence of inter-layer coupling the chemical
potential is chosen so that $\langle n \rangle=0.8$. Lastly, we consider
the case where $\Delta_{\pm}=15$ meV for both bands. 

The resulting spectra for the bonding and anti-bonding bands, as well as
the mixing term for $A_{1g}$ are shown in Fig. 3. Here again we have used
the effective mass approximation for the expansion parameters of Eq. 7
for illustration\cite{efm}. Considering the $B_{1g}$ and $B_{2g}$
channels, which are simply additive, it can be inferred from experimental
result which shows only one peak in these channels that either the
energy gap is nearly identical both in symmetry and magnitude or alternatively
that the Raman response is given predominantly by one band. Otherwise
two distinct peaks would appear (if gaps have different symmetry and/or
magnitude) and identical power-law behavior for each channel
would be seen (if the gaps
were of different symmetry but same magnitude)\cite{cave}. Since the band
splitting is relatively small in Y 1:2:3 and even smaller in Bi 2:2:1:2,
the second possibility in our opinion is unlikely. Therefore, the energy
gap on each band must be similar on energy scales determined by the amount
of inelastic scattering which would smear a double-peak feature. Again we
note that the sharp nature of the $B_{1g}$ peak would imply that its
position and shape would be most sensitive to the location of the Van Hove
points for each band and the maximal value of the energy gaps\cite{slakey,alt}.

Due to the small bi-layer splitting the mixing term for $A_{1g}$ is small, 
as shown in Fig. (3). A very small peak at roughly $2\Delta_{max}$ 
(not divergent) is seen due to the odd
combination of interlayer charge transfer, as suggested by Krantz and
Cardona but is strongly surpressed. Since adding electron correlations
supresses the bi-layer splitting even further\cite{oka2}
the mixing term is only of minor importance and
the resulting spectra can be well approximated as the sum of the 
contributions from the two individual bands. In this way, a consistent
picture can emerge since experimentally the electronic Raman spectra do
not differ substantially in the bi-layer or single-layer materials
\cite{tpd,irwin,exp}.

\section{Impurity and Spin Fluctuation Scattering}

The previous considerations were limited to the case of clean superconductors
in which light scattering only occurs via breaking of Cooper pairs. Since the
limit of ${\bf q \rightarrow 0}$ was taken, as a consequence no scattering
remains above $T_{c}$ due to phase space constraints. Impurity and/or 
spin fluctuation scattering must therefore be taken into account in order
to attempt to model the full spectra for a wider range of temperatures
and frequencies. In closing, this section is devoted to including such
scattering processes. The following discussion is mostly qualitative in 
nature.  An earlier account of this work is provided in Ref. \cite{imp1}. 

Isotropic impurity scattering can be easily incorporated using a T-matrix
approach. This simplified approach [recent studies have shown that impurity 
scattering in a strongly correlated material can have a non-trivial 
$k-$dependence\cite{hanke}] does not mix symmetry channels
and therefore previous considerations can be applied. 
The scattering is then characterized
the by the phase shift $\delta$, $c=\cot(\delta)$, and by the impurity
concentration $n_{i}$, $\Gamma=n_{i}/\pi N_{F}$. Since we have seen that
the shape of the Fermi surface has little effect on the low energy part
of the cross section, we for simplicity take an infinite band and a nearly
cylindrical Fermi surface [see Ref. \cite{tpdde} for details]. Here we
also consider only unitary scattering $c=0$, and small impurity concentrations
$\Gamma/\Delta_{max}=0.01$. Other choices for the parameters are considered
in a forthcoming publication\cite{forth}. While certainly a simplication of
real systems, the salient features can be explored with some generality.

In order to incorporate spin fluctuation scattering, we follow the work of
Ref. \cite{sean}, who considered spin fluctuation scattering in a $d_{x^{2}-
y^{2}}$ superconductor in the context of the infrared conductivity using
a 2-D Hubbard model. There spin fluctuation scattering was incorporated in the 
imaginary part of the self energy while vertex corrections and the real part
of the self energy were neglected. We follow the same approach here
and use their interpolation result for the frequency dependent inelastic 
scattering rate, which was found 
at low temperatures to vary as
$\omega^{3}$ for frequencies below three times the maximum gap
energy and then linearly thereafter:
\begin{equation}
\Gamma_{s.f}(\omega,T << T_{c})/3\Delta_{max}=0.7
\cases{\ \ \ \ \left(\omega/3\Delta_{max}\right)^{3} & 
for $\omega < 3\Delta_{max}$, \cr
\omega/3\Delta_{max} & otherwise},
\end{equation}
which was obtained for $T=0.1T_{c}, \langle n \rangle=0.85, U=2$ in a
model with only nearest neighbor hopping.
The scattering rate is approximated to be isotropic.

The additional scattering mechanisms modify the spectra in two important
regards - the low frequency behavior, caused by impurity scattering, and
the high frequency behavior caused by spin fluctuation scattering. This
is shown in Fig. (4) for the $B_{1g}$ and $B_{2g}$ channels. 

The impurity modifications on the density of states is reflected in the
low frequency part of the spectra. Here the foot in the density of states
causes the low frequency part of the spectra in each channel to rise linearly
in frequency as it does in the normal state. This holds for the $B_{1g}$
channel as well, where a crossover is seen at a frequency $\omega^{*}
\sim \sqrt{\Gamma \Delta_{max}}$ which separates linear behavior below
and cubic behavior above as seen in Fig. 4. 
This is the opposite of what one would expect
if the energy gap were of anisotropic $s-$wave symmetry, since there
spectral weight would be shifted to {\it higher} frequencies due to an
opening of a gap\cite{bork,imp1}. The only changes at higher frequencies
comes from a smearing of the singularities at twice the gap edge.

Spin fluctuation scattering is relatively
inoperative at low frequencies and thus
does not produce any noticeable effects. However, the approximately
linear in frequency behavior at large frequencies, as well as the
value of the scattering rate at twice the gap edge modify the spectra
at larger frequency shifts. The scattering at twice the gap edge effectively
smears out any singularities due to the energy gap (or to the Van Hove). 
Further, a flat featureless background at larger frequencies is obtained.
Similar results have been obtained by Jiang and Carbotte using a different
model\cite{imp2}.

\section{Conclusions}

We have seen that electronic Raman scattering contains a wealth of information
concerning the nature of superconductivity in unconventional superconductors
which can be extracted from symmtery considerations alone\cite{phonons}. 
When appied to
the experimental data for ${\it optimally}$ doped cuprate materials,
good agreement can be obtained for $d_{x^{2}-y^{2}}$ pairing in single band 
materials, and nearly indentical gaps of $d_{x^{2}-y^{2}}$ symmetry 
for bi-layer systems. The analysis is based on the prominent behavior of
the spectra at low frequencies and the approximate position of the 
maxima for each channel. 

However, further work is needed in order to understand the properties of
the Raman spectra {\it away} from optimal doping\cite{exp,slakey,alt,pri}. 
It is suggested that the $B_{1g}$ peak may be the most sensitive to doping
due to the role of the Van Hove singularity. However in order to understand
the effects of doping in more detail, a particular model of
superconductivity is needed, as is a more sophisticated approach for
handling the spin degrees of freedom.

\section{Acknowledgements}

This work in part is done in collaboration with Drs. D. Einzel, A. Virosztek,
and A. Zawadowski, and was supported by NSF Grant No. DMR 95-28535.

\begin{table}
\centerline{TABLE I}
\centerline{Summary of Raman response for various pair state candidates for 
clean, tetragonal superconductors.$^{*}$} 
\begin{tabular}{|l|l|l l|l l|l l|} \hline\hline
{$\Delta({\bf k})$} & {$N(\omega \rightarrow 0)$} &
\multicolumn{2}{|c|}{$B_{1g}$}   &
\multicolumn{2}{|c|}{$B_{2g}$}   & \multicolumn{2}{|c|}{$A_{1g}$}  
\\ 
& & {$\chi^{\prime\prime}(\omega \rightarrow 0)$} & ${\omega_{peak}\over{\Delta_{max}}}$ & {$\chi^{\prime\prime}(\omega \rightarrow 0)$} & ${\omega_{peak}\over{\Delta_{max}}}$ & {$\chi^{\prime\prime}(\omega \rightarrow 0)$} & ${\omega_{peak}\over{\Delta_{max}}}$\\
\hline\hline
isotropic $s-$wave & $\Theta(\omega-\Delta)$& $\Theta(\omega-2\Delta)$ & 2 & $\Theta(\omega-2\Delta)$ & 2 & 
$\Theta(\omega-2\Delta)$& 2\\  \hline
$d_{x^{2}-y^{2}}$ & $\omega$ & $\omega^{3}$ & 2 & $\omega$ & $\sim 1.7$ & $\omega$ & $\sim 1.2 $ 
\\ \hline
$s+id_{x^{2}-y^{2}}$ & $\Theta(\omega-\Delta_{s})$ & $\Theta(\omega-2\Delta_{s})$ & 2 & 
$\Theta(\omega-2\Delta_{s})$ & $2\Delta_{s}/\Delta_{d}$ & $\Theta(\omega-2\Delta_{s})$ & $2\Delta_{s}/\Delta_{d} $ 
\\ \hline
$d_{xy}$ & $\omega$ & $\omega$ & $\sim 1.7$ & $\omega^{3}$ & $2$ & $\omega$ & $\sim 1.2 $
\\ \hline
$(d_{x^{2}-y^{2}})^{m}$ & $\omega^{1/m}$ & $\omega^{3/m}$ & 2 & $\omega^{1/m}$ & $< 1.7$ & $\omega^{1/m}$ & $\le 1.2 $
\\ \hline
$g-$wave$^{\dagger}$ & & & & & & & \\
$\sim \cos(k_{x}a)\cos(k_{y}a)$ & $\omega$ & $\omega$ & $2$ & $\omega$ & $0.6$ & $\omega$ & 
$ 0.6 $
\\ \hline
extended $s-$wave$^{\dagger\dagger}$ & & & & & & & \\
$\sim \cos(k_{x}a)+\cos(k_{y}a)$ & $\omega$ & $\omega$ & $\sim 1.7$ & $\omega$ & $2$ & $\omega$ & $\sim 1.2 $
\\ \hline \hline
experimental results & & & & & & & \\
for optimal doping & $\omega$ & $\omega^{3}$ & 2 & $\omega$ & $1.5\sim 1.8$ & $\omega$ & $1.0 \sim 1.2 $
\\ \hline
\end{tabular}
$^{*}$ Here phase is undetermined and thus $\mid d_{x^{2}-y^{2}}\mid$ would
yield the same as $d_{x^{2}-y^{2}}$. Moreover, orthorhombic distortions 
will mix $B_{1g}$ and $A_{1g}$ channels.\\ 
$^{\dagger}$ Parameters chosen to split node at $45^{o}$ into two 
modes at $45^{o} \pm 5^{o}$ and a subsidiary 
maximum equal to 20 percent of gap maxima. \\
$^{\dagger\dagger}$ Parameters chosen such that $\langle n \rangle=0.8$.
\end{table}
\begin{figure}[t]
\epsfxsize=2.5in
\epsfysize=4.0in
\epsffile{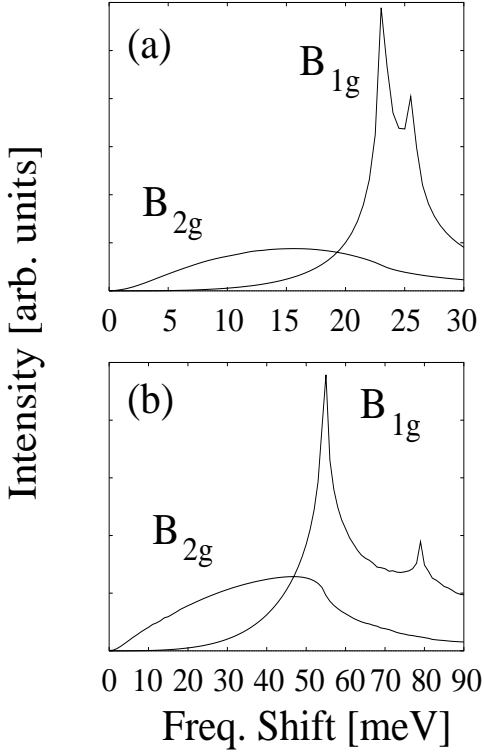}
\caption{$B_{1g}$ and $B_{2g}$ Raman responses for a single band 
plotted for parameters sets for (a) La 2:1:4 and (b) Y 1:2:3.
Here a filling $\langle n \rangle =0.8$ and $t=100$ meV, and 
$\Delta(k)=\Delta_{0}[\cos(k_{x}a)-\cos(k_{y}a)]$ are used for both figures. 
The other parameters are: (a) $t^{\prime}/t=0.16, t^{\prime\prime}=0, 
\Delta_{0}=6$meV; (b) $t^{\prime}/t=0.2, 
t^{\prime\prime}/t=0.25, \Delta_{0}=15$ meV.}
\end{figure}
\newpage
\begin{figure}[t]
\vskip -2.0cm
\epsfxsize=2.5in
\epsfysize=3.5in
\epsffile{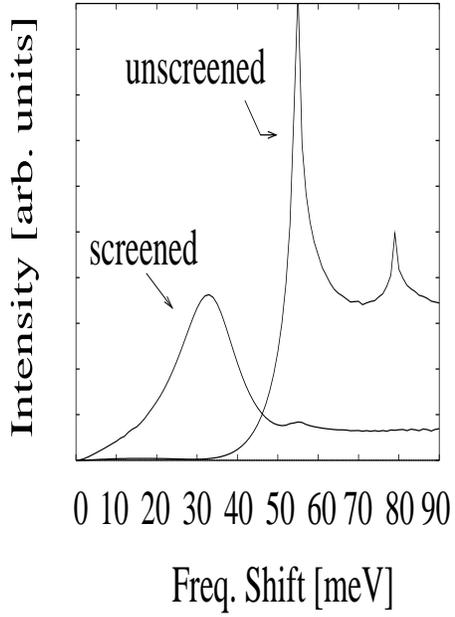}
\caption{$A_{1g}$ response calculated for a single band 
with and without Coulomb screening for the Y 1:2:3
parameters.}
\end{figure}
\begin{figure}[t]
\vskip 0.5cm
\epsfxsize=2.5in
\epsfysize=3.5in
\epsffile{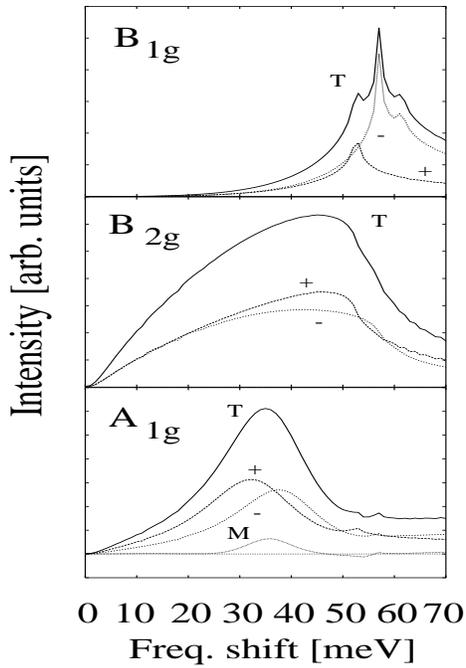}
\caption{Raman response calculated for Y 1:2:3 bi-layer for various channels
as indicated. The $+ (-)$ indicates the bonding (anti-bonding) 
band, respectively, the symbol $M$ is the mixing term [Eq.
(14) which only contributes for the $A_{1g}$ channel], and the total response
is indicated by the symbol $T$.}
\end{figure}
\newpage
\begin{figure}[t]
\epsfxsize=3.5in
\epsfysize=4.5in
\epsffile{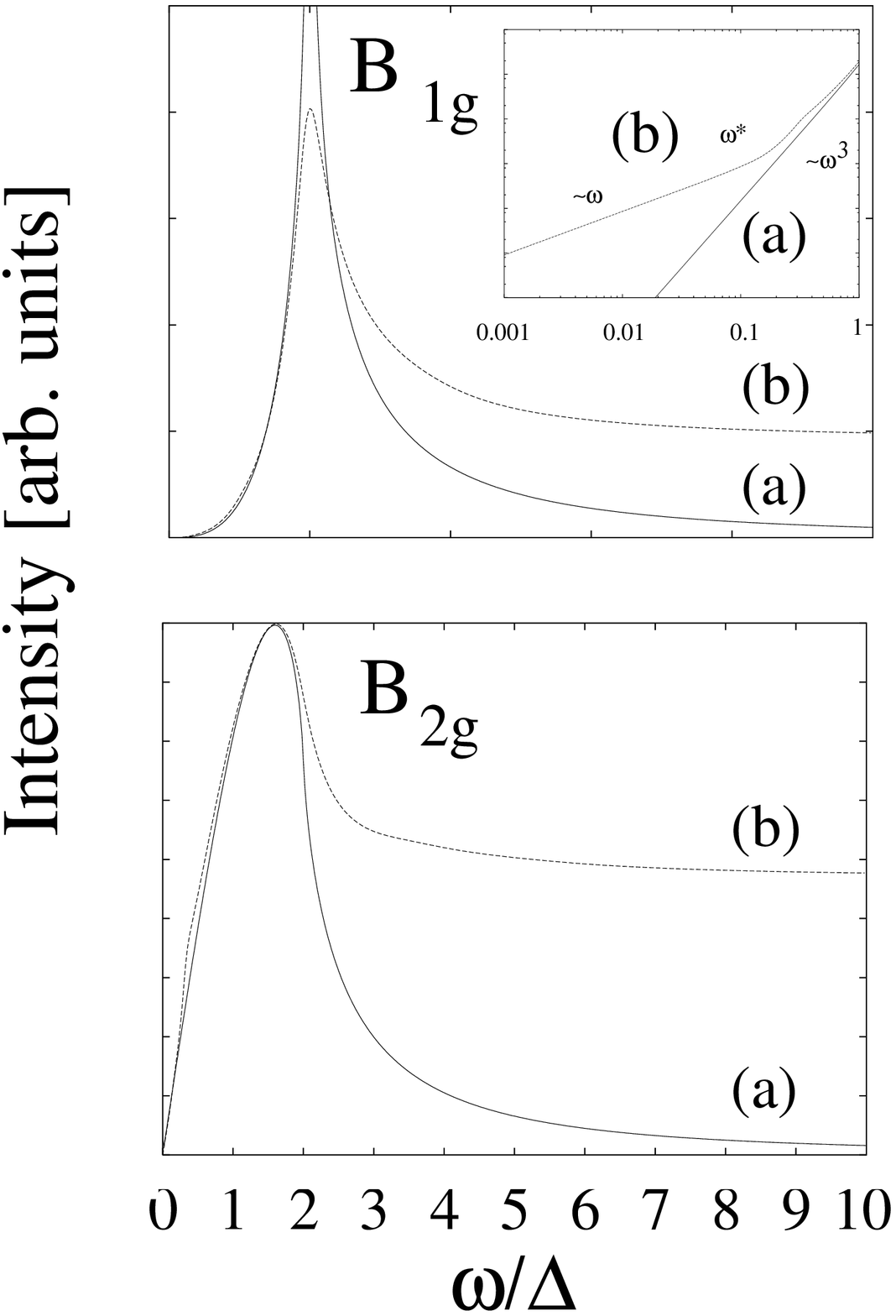}
\caption{Raman response calculated for $B_{1g}$ and $B_{2g}$ channels
in the absence (a) and presence (b) of impurity and spin fluctuation 
scattering. Inset shows the low frequency crossover of the $B_{1g}$ channel.
Parameters used are described in text.}
\end{figure}

\begin{thebibliography}{99}
\bibitem{cave}
The power-law behavior would not be changed if one of the gaps was the 
same magnitude as the other but completely isotropic $s$-wave. However,
then all channels would show a sharply defined peak at $2\Delta_{max}$.
This is in disagreement with experiment and appears to be unlikely.
\end{thebibliography}
\end{document}